\documentclass[aps,pra,nofootinbib,preprint,amsmath,amssymb,floatfix]{revtex4-2}
\usepackage{color}
\usepackage{graphicx}

\begin{document}

\newcommand{\tc}{\textcolor}
\newcommand{\g}{blue}
\newcommand{\ve}{\varepsilon}
\title{ Possible expansion of blood vessels by means of the electrostrictive effect
 \emph{}}         

\author{ Iver Brevik }      

\affiliation{Department of Energy and Process Engineering, Norwegian University of Science and Technology, N-7491 Trondheim, Norway}

\date{\today}          

\begin{abstract}
In cases when  it is desirable to transport medication through blood vessels, especially when dealing with  brain cancer being confronted with   the narrow arteries in the brain, the blood-brain barrier makes the medical  treatment difficult. There is a need of expanding the diameters of the arteries in order to facilitate the transport of medicaments.  Recent research has pointed to various ways to improve this situation; in particular, the use  ultrasound acting on microbubbles in the blood stream has turned out to be a promising option. Here, a different possibility of enlarging the diameters of arteries is discussed,   namely to exploit the electrostrictive pressure produced by internal strong, ultrashort and   repetitive laser pulses. Each pulse will at first  give rise to inward directed optical forces, and  once the pulse terminates there will be a hydrodynamical bouncing flow in the outward radial direction giving an outward impulse to the vessel wall. In the absence of friction a symmetric oscillation picture emerges. Clearly, a supply of repetitive pulses will be needed (at parametric resonance)   to make the effect appreciable. The effect has to our knowledge not  been discussed before. We give an approximate  optical and hydrodynamical theory of it. The calculations indicate promising results for the wall pressure, although experimental work is desirable to show whether the idea can be useful in practice. Our calculation is made from a  general physical perspective, not necessarily linked to medical applications.

\end{abstract}
\maketitle

\section{Introduction}

As is known,  symmetry properties often play a dominant role in     fluid mechanical systems; cf., for instance the collection of papers discussing such themes in  \cite{specialissue}.   Here, we will consider a situation which combines  classical laser physics with hydromechanics: Assume that a long liquid-filled circular tube is exposed to one or more  longitudinal short laser pulses. If the fluid is compressive, each pulse  will give rise to a radial electrostrictive forces in the inward direction. When the pulse has left, the fluid rebounds, producing in turn an outward hydrodynamic pressure  on the wall. Such a force may be of physical interest. If the fluid is nonviscous, there results  a stationary symmetric oscillatory wave situation.

The physical motivation for the present study came actually  from a well known problem in medical technology. The human brain has a filter,  that protects it as well as the rest of our nervous system from foreign elements that can damage the tissue. This is the blood-brain barrier, and is normally a useful and vital organ. A problem arises, however, in the case of diseases when one wants to make medication penetrate the barrier. This filter problem is accentuated in the case of brain tumours. Recent research has indicated that there are promising measures that can be taken to overcome the problem such as to inject microbubbles into the blood, and thereafter expose the brain to focused ultrasound; cf., for instance, Refs.~\cite{olsman21} and \cite{poon22}. This "Acoustic Cluster Therapy" (ACT) as it is called, is an ingenious method; the ultrasonic waves transfer energy to the microbubbles, which in turn merge into a single large microbubble effectively opening the blood-brain barrier.

Some general remarks: the blood-brain-barrier (BBB), as a semi-permeable membranous barrier, is located at the interface between the blood and the cerebral tissue. It is composed of a complex system of  endothelial  and perivascular mast cells, and is responsible for controlling the exchanges between two compartments allowing only certain molecules or ions to pass through. In the process of drug delivery, only lipid soluble molecules with a low weight and of positive charge can cross the BBB.  Other molecules require  carrier-mediated transport systems. Some strategies consist of the injection of therapeuthic proteins directly into the cerebrospinal fluid. A promising approach seems to be the use of lipid- and polymer-based nanosized particles that assure a controlled release of their cargo by protecting loaded drugs from being metabolized. A useful review article on crossing the blood-brain barrier can be found in Ref.~\cite{bellettato18}.

Among other methods, we mention how a magnetically driven soft continuum microrobot can be used for intravascular operations in microscale \cite{liu22}. Whereas conventional continuum robots have the miniaturization challenge, this paper reviews a microscale soft continuum robot with steering and locomotion capabilities based on magnetic field actuation. In  Ref.~\cite{song21}, a new approach is discussed about how the combination of a FET biosensor with a rolled-up microtube can be used to develop a microfluidic diagnostic biosensor. Finally, in Ref.~{\cite {lipsman18}, the BBB is discussed in connection with patients having Alzheimer's disease; the use of MR-guided focused ultrasound turned out to be a promising avenue experimentally.

\bigskip

In this paper we will discuss a different approach to the enlargement problem. As already mentioned, to our knowledge it has not been considered before.  Assume that an ultrashort strong laser pulse   is sent longitudinally in the $z$ direction through the blood vessel. As blood is a compressible liquid, there will be electrostrictive radial forces compressing the liquid inward, as long as the pulse lasts. From a hydrodynamic point of view, this compressive event can be regarded to be instantaneous. When the pulse has left, the hydrodynamic response causes the fluid to move radially inwards, and build up a maximum pressure around the symmetry axis.  Assuming as an example  that the pulse has a strong power of 3~kW, the excess hydrodynamic pressure on the axis will be quite large, about one atmosphere. After this, a hydrodynamic rebound occurs, causing the fluid to move radially outwards and transfer an outward momentum to the vessel wall. The liquid then flows back, and becomes compressed near  the centerline $r=0$ again. At first, we will neglect the shear viscosity, so that a hydrodynamic stationary wave pattern becomes established. To make the effect big enough in practice, it is clear that a large number $N$  of successive laser pulses per second will be  needed. This is actually an example of  parametric resonance:  every new pulse is imaged to be sent through the system during the phase  when the flow is running inwards. If one as the next point takes into account  viscosity, it turns out that the viscous damping will not have time to make any appreciable influence on the first rapid hydrodynamic oscillations at all. For longer time spans, the viscosity will naturally be important.

Note: we have here assumed that the compressed fluid is temporarily prevented from leaking out in the longitudinal direction. That is, there must be an effective blockade against longitudinal fluid motion at the instant just after the onset of the pulse. This seems to be a reasonable assumption for these short times,  in view if the many sharp turns of the artery system. For longer times, of course, the fluid is free to move longitudinally.

The method described is conceptually simple, and our calculations below indicate that the electrostrictive forces  may  give rise to an effect of noticeable magnitude. To describe the physics of the opto-electromagnetic process accurately, would however seem  difficult. To get a realistic knowledge about the usefulness of the method in practice, one needs  experimental investigations, preferably at larger dimensions than those envisaged here.

For readers interested in further works about the electrostrictive effect considered from various perspectives, we list a number of papers (not meant to be exhaustive):  Refs.~\cite{nelson79,brevik79,rasaiah81,carnie82,rasaiah82,brevik82,zimmerli99,hallanger05,birkeland08,brasselet08,delville09,wunenburger11,
ellingsen11,ellingsen12,astrath14,astrath22,astrath23}.

\section{The electrostrictive force density }

The basic expression for the electrostrictive electromagnetic force density is \cite{stratton41,jackson99}
\begin{equation}
{\bf f} = \frac{1}{2}\varepsilon_0 {\bf \nabla}
\left( E^2\rho_m \frac{d\varepsilon}{d\rho_m}\right), \label{1}
\end{equation}
where $\varepsilon = n^2$ with $n$ the refractive index, and $\rho_m$ is the density of the fluid. We assume the medium to be nonmagnetic. The electric field $E$ refers to the rms-value. Strictly speaking, the equation above refers to  as  a nonpolar  medium, so that $\varepsilon$ is a function of $\rho_m$ only, but the equation is approximately valid for many other media also. We will moreover make use of the Clausius-Mossotti relation.  This implies, that if we in general writes the force density as
\begin{equation}
{\bf f} = {\bf\nabla}\chi, \label{2}
\end{equation}
we get the following  explicit expression for the scalar $\chi$,
\begin{equation}
\chi = \frac{1}{6}\varepsilon_0 E^2 (\varepsilon-1)(\varepsilon +2). \label{3}
\end{equation}

Consider now the intensity of the incident laser pulse, which is taken to be  constant during the time $t_0$ the  pulse lasts.  With cylindrical coordinates ${\bf r}=(\rho, \theta,z)$
 we shall assume the form
 \begin{equation}
 E^2({\bf r},t) = E^2({\bf r})T(t), \label{4}
 \end{equation}
where $T(t)$ is the temporal function of the pulse,
\begin{equation}
T(t)= \Theta(t)-\Theta(t-t_0), \label{5}
\end{equation}
$\Theta(t)$ being the step function, equal to +1 when $t>0$ and 0 when $t<0$.  For the spatial variation of the quadratic electric field we adopt the Gaussian form
\begin{equation}
E^2({\bf r})= \frac{2P}{\pi \varepsilon_0ncw^2(z)}\exp \left[ -\frac{2\rho^2}{w^2(z)}\right], \label{6}
\end{equation}
where $P$ is the power of the pulse, and $w(z)$ is the beam radius
\begin{equation}
w(z)= w_0\sqrt{1+z^2/l_R^2}, \label{7}
\end{equation}
with $w_0$ the minimum radius at the waist.
The length $l_R = \pi w_0^2/\lambda$ is called the Rayleigh length, where   $\lambda$ is the wavelength in the medium. We will focus on the region near the center of the beam (taken to be $z=0$), assuming $z/l_R \ll 1$, so that $w \approx w_0$.
It means neglect of the convergence of the beam, and  complies with the restriction $\lambda/w_0 \ll 1$.

The approximate form of the radially dependent part $\chi(\rho)$ of the scalar $\chi(\rho,t)$ thus becomes
\begin{equation}
\chi(\rho)= \chi(0)\exp \left[ -\frac{2\rho^2}{w_0^2}\right], \label{8}
\end{equation}
where the value on the centerline
\begin{equation}
\chi(0)= \frac{(\varepsilon-1)(\varepsilon+2)P}{3\pi nc w_0^2} \label{9}
\end{equation}
is a constant. The full, time-dependent, electrostrictive force density is
\begin{equation}
{\bf f}(\rho,t)= \chi(0)T(t)\frac{d}{d\rho}\left[e^{-2\rho^2/w_0^2}\right]\hat{\bf \rho }. \label{10}
\end{equation}
The expression is negative, corresponding to a compressive radial force. The force is constant in time, as long as the pulse lasts.

\section{Hydrodynamic analysis of the fluid motion. The build-up phase}

There are clearly two widely separated time scales in this problem. The electrodynamic scale is determined by the light velocity $c$ (for definiteness taken in vacuum), while the hydrodynamic scale is determined by the sound velocity $u$. Hydrodynamically, the pulse of  duration $t_0$ can be regarded as an instantaneous flash, transferring an impulse to the fluid. We will first write down the general governing equations for the fluid motion, neglecting the viscosity. Let $\bf V$ be the fluid velocity, and start from the   equation of continuity,
\begin{equation}
\partial_t\rho_m +{\bf \nabla}\cdot (\rho_m{\bf V})=0. \label{11}
\end{equation}
A linear expansion to first order, $\rho_m= \rho_{m0}+ \rho_m', \, p=p_0+p'$, with $\rho_m' \ll \rho_{m0}, \, p' \ll p_0$, yields then the linearized version
\begin{equation}
\partial_t\rho_m' + \rho_{m0}{\bf \nabla}\cdot {\bf V}=0. \label{12}
\end{equation}
Next, consider the Euler equation when the nonlinear convective term is neglected,
\begin{equation}
\rho_m\partial_t{\bf V}= -{\bf \nabla}p+{\nabla \chi}. \label{13}
\end{equation}
 In linearized form it reads
 \begin{equation}
 \rho_{m0}\partial_t{\bf V}=-{\bf \nabla}p'+{\bf \nabla}\chi. \label{14}
 \end{equation}
 From this the sound velocity $u$ follows naturally via the equation
\begin{equation}
 p' = \left( \frac{\partial p}{\partial \rho_{m0}}\right)_S\rho' = u^2\rho'. \label{14}
 \end{equation}
 This equation is used to replace the $\rho_m'$ term in the continuity equation  (\ref{12}).
[A note on the time scales: after imposition of the pulse, the time required to reach the state of maximum pressure is of order $w_0/u$. With $w_0$ a few $\mu$m and $u=1500~$ms$^{-1}$, the building-up time becomes of order ns. This short time makes it natural to apply the subscript S (constant entropy) in Eq.~(\ref{14}).]

As the fluid is so far assumed nonviscous, we can introduce the velocity potential $\Phi$,
\begin{equation}
{\bf V} = {\bf\nabla}\Phi. \label{15}
\end{equation}
Introducing this into the Euler equation (\ref{13}) we obtain, after removal of the nabla operator,
\begin{equation}
p'= -\rho_{m0}\partial_t\Phi +\chi. \label{16}
\end{equation}
Differentiating this with respect to $t$ and comparing with the continuity equation, we obtain the general governing equation for $\Phi$,
\begin{equation}
\nabla^2\Phi(\rho,t)-\frac{1}{u^2}\partial_t^2\Phi(\rho,t)= -\frac{1}{\rho_{m}u^2}\partial_t\chi (\rho,t), \label{17}
\end{equation}
in which we have replaced $\rho_{m0}$ by $\rho_m$.

Consider now the case where the time function $T(t)$ has the form of Eq.~(\ref{5}), from which we derive
\begin{equation}
\partial_t\chi(\rho,t) = \chi(\rho)[\delta(t)-\delta(t-t_0)]. \label{18}
\end{equation}
 Equation (\ref{17}) can be solved using Green function methods. We shall do that here, although, as a simplifying element, we shall assume that there is an infinite fluid around the laser beam. To include the finite width of the blood vessel would be quite complicated, and not justified by the approximate calculation here given. Our
  simplified equation for the Green function becomes thus
 \begin{equation}
 (\nabla^2- u^{-2} \partial_t^2)G({\bf r,r'},t,t')
  = \delta({\bf{r'-r}})\delta(t'-t), \label{19}
 \end{equation}
 which has the solution
 \begin{equation}
 G({\bf {r,r'}},t,t') =
  -\frac{1}{4\pi |{\bf{r'-r}}|}
  \delta \left( t'-t + \frac{|{\bf{r'-r}}|}{u} \right). \label{20}
 \end{equation}
 Then the velocity potential can be written as
\begin{equation}
\Phi(\rho,t)= -\frac{1}{\rho_m u^2}\int_{-\infty}^t dt'\int d^3 r'G({\bf r},{\bf r'},t,t')\partial_t \chi(r',t'). \label{21}
\end{equation}
It is now convenient to introduce a nondimensional length $\xi$ and a nondimensional time $\tau$,
\begin{equation}
\xi= \frac{\sqrt{2}\,\rho}{w_0}, \quad \tau =
\frac{{\sqrt 2}\,ut}{w_0}. \label{23}
\end{equation}
Under the present conditions, they are both of order unity. Then, the velocity potential can be written as \cite{ellingsen11}
\begin{equation}
\Phi(\rho,t)= \frac{\chi(0)w_0}{\sqrt{2}\,u\rho_m}\left[ \tilde{\Phi}(\xi, t)-\tilde{\Phi}(\xi, \tau-\tau_0)\right], \label{24}
\end{equation}
\begin{equation}
\tilde{\Phi}(\xi, \tau)= \Theta(\tau)e^{-\xi^2 - \tau^2}\int_0^\tau dx I_0(2\xi \sqrt{\tau^2-x^2})e^{x^2}, \label{25}
\end{equation}
where $I_0$ is the modified Bessel function of the first kind of order zero. The pressure becomes
\begin{equation}
p(\xi,\tau)= \chi(0)[P(\xi, \tau)-P(\xi, \tau-\tau_0)], \label{26}
\end{equation}
\begin{equation}
P(\xi, \tau)= 2\tau e^{-\xi^2}\Theta(\tau) \int_0^\tau \frac{xdxe^{-x^2}}{\sqrt{\tau^2-x^2}} \left[ I_0(2\xi x)-\frac{\xi}{x}I_1(2\xi x)\right]. \label{27}
\end{equation}
Of special interest is the build-up of  pressure on the symmetry axis, $\xi=0$. Then, the formulas above  reduce to
\begin{equation}
\tilde{\Phi}(0, \tau)= \Theta(\tau)F(\tau), \label{28}
\end{equation}
where
\begin{equation}
F(x)= e^{-x^2}\int_0^xe^{t^2}dt \label{29}
\end{equation}
is Dawson's integral, and
\begin{equation}
p(0,\tau)= 2\chi(0)[\Theta(\tau)\tau F(\tau)-\Theta (\tau-\tau_0)(\tau-\tau_0)F(\tau-\tau_0)]. \label{30}
\end{equation}
The numerically calculated Fig.~4 in Ref.~\cite{ellingsen11} shows for typical input data how the pressure nicely builds on the axis, almost linearly, until a maximum somewhat in excess of $\chi(0)$  is reached around the time it takes for sound to traverse the laser beam. For practical purposes we may say that
\begin{equation}
p_{\rm max} = \alpha \chi(0), \quad \alpha \approx 1.25. \label{31}
\end{equation}
After this maximum the pressure decreases slowly as long as the pulse lasts, and for $\tau > \tau_0$ it decreases quickly.

Our next task will be to calculate the elastic energy in the fluid, imparted by the pulse. Only the first, build-up, phase will then be of interest. We take the pulse to start at $t=0$, and follow it until maximum pressure is established. We start from the velocity potential given in Eq.~(\ref{25}), but make the simplifying assumption that the argument $2\xi \sqrt{\tau^2-x^2}$ in  $I_0$ is small. As this Bessel function increases only slowly for small arguments, $ I_0(z) \approx 1+\frac{1}{4}z^2$,    this restriction is not very strong. In the lower limit $x=0$ the argument is $2\xi\tau$, whereas in the upper limit $x=\tau$  the argument is zero. One may estimate that the main contribution to the integral occurs when $ut \sim w_0/2$, i.e., $\xi \sim 1/\sqrt{2}$, and at the position $\rho \sim w_0/2$, i.e., $\xi \sim 1/\sqrt{2}$, what corresponds to $2\xi \tau \sim 1$ at most. To the present accuracy, the approximation is acceptable.

In this way we obtain an expression for the velocity potential in which Dawson's integral is the main ingredient,
\begin{equation}
\tilde{\Phi}(\xi,\tau)= e^{-\xi^2}F(\tau), \label{32}
\end{equation}
what gives the following expression for the dimensional radial fluid velocity,
\begin{equation}
V(\rho,t)= \frac{d\Phi (\rho,t)}{d\rho} = \frac{\chi(0)}{u \rho_m}F(\tau)\frac{d}{d\xi}e^{-\xi^2}. \label{33}
\end{equation}
Similarly, the radial component of the electrostrictive force density becomes
\begin{equation}
f(\rho)= \chi(0)\frac{\sqrt{2}}{w_0}\frac{d}{d\xi}e^{-\xi^2}. \label{34}
\end{equation}
It is worth noticing that whereas $V(\rho,t)$ depends on time, $f(\rho)$ does not. The latter behavior is related to the constancy of the pulse as long as it acts.

The produced elastic energy $E_{\rm elastic}$ (per unit length) in the fluid is equal to  the work done by the inward directed electrostrictive forces,
\begin{equation}
E_{\rm elastic} = \int 2\pi \rho d\rho f(\rho)V(\rho, t)dt. \label{35}
\end{equation}
Making use of Eqs.~(\ref{33}) and (\ref{34}) we obtain after some calculation
\begin{equation}
E_{\rm elastic}= \frac{\pi w_0^2}{2\rho_mu^2}\,\chi^2(0)\int_0^{\tau_{\rm max}}F(\tau)d\tau, \label{36}
\end{equation}
where $p_{\rm max}$ is given in Eq.~(\ref{31}).

We now turn to some numerical estimates. Our main interest is the small arteries; as informed in the review article \cite{camasao21}, these arteries may have an outer diameter of about 1~mm. With a wall thickness of about $200~\mu$m,   this gives an inner radius $a$ of a few hundred $\mu$m. For calculational convenience we will choose the low value $100~\mu$m in the following. This will also enhance the pressure effect. The refractive index for blood is 1.37; the density is $10^3~$kg~m$^{-3} $ as for water, and  for the power $P$ of the pulse we  adopt the same value 3 kW as used by Ashkin and Dziedzic in their classic radiation pressure experiment on water \cite{ashkin73} (this value reported to lie below the region of nonlinearity). For the beam radius at the waist, we choose $ 2~\mu$m. To summarize,
\begin{equation}
a=100~\mu{\rm m},  \quad n=1.37, \quad  P= 3~{\rm kW}, \quad w_0= 2~\mu{\rm m}. \label{37}
\end{equation}
A remark on the nondimensional time $\tau$: as defined in Eq.~(\ref{23}) this parameter will take a  convenient value of order unity when $t= w_0/u$, which is the hydrodynamically important scale. This follows directly from $\tau= \sqrt{2}t/(w_0/u) \sim 1$ at that instant.

We can now calculate, from Eq.~(\ref{9}),
\begin{equation}
\chi(0)= 4.9\times 10^5~\rm{Pa}, \label{37}
\end{equation}
what is  quite a high pressure, about 5 atmospheres. The elastic energy evaluates to
\begin{equation}
E_{\rm elastic}= 6.6\times 10^{-10}\int_0^{\tau_{\rm max}}F(\tau) d\tau. \label{38}
\end{equation}
The function $F(\tau)$ has a maximum 0.541 at $\tau = 0.924$. The integral (\ref{38}) is divergent for high arguments, but we have made a simple numerical evaluation of it assuming the upper limit $\tau_{\rm max}=2$, which seems to be  reasonable.  The result was approximately 0.8, and so we obtain finally
\begin{equation}
E_{\rm elastic}= 5.3\times 10^{-10}~{\rm Jm}^{-1}. \label{39}
\end{equation}
This is thus the energy imparted to the fluid from one single pulse.

\section{Hydrodynamic analysis: the nonviscous oscillating phase}

We now analyse the fluid motion after the pulse has left. To begin with, defined hereafter as  $t=0$ the fluid is at rest at maximum pressure, and for  $t>0$ it rebounds radially outwards, giving rise to a radial oscillating wave pattern. As long as viscosity is ignored, the pattern will be stationary. When reflected against the outer wall, the fluid imparts an outward  force on it. This is the force we intend to calculate. At first, we consider the influence from one single pulse only.

The governing equation for the velocity potential is now Eq.~(\ref{17}) with the right hand side equal to zero,
\begin{equation}
\left( \partial_\rho^2+\frac{1}{\rho}\partial_\rho -\frac{1}{u^2}\partial_t^2\right)\Phi(\rho,t)= 0. \label{41}
\end{equation}
We will write the solution as
\begin{equation}
\Phi(\rho,t)= -A J_0(k\rho)\sin \omega t, \quad k=\omega/u, \label{41a}
\end{equation}
with $J_0$ the ordinary Bessel function of order zero and $A$ a positive constant. The radial velocity, positive to begin with, is thus
\begin{equation}
V(\rho,t)=\partial_\rho \Phi(\rho,t)= kAJ_1(k\rho)\sin \omega t. \label{42}
\end{equation}
The kinematic condition $J_1(ka)=0$ at the wall (the radial movement of the wall negligible in this context) determines the allowable values of the wave number $k$. We shall consider the lowest mode only,
\begin{equation}
ka = 3.83, \label{43}
\end{equation}
which corresponds to the wavelength $\lambda = 2\pi/k = 1.64 \,a$ of the fluid oscillations. The angular frequency becomes
\begin{equation}
\omega = \frac{5.75}{a}\times 10^3~{\rm rad\, s}^{-1}, \label{44}
\end{equation}
which means a frequency of $\omega/2\pi \approx 10\,$ MHz when $a= 100~\mu$m.

We now calculate the kinetic energy $E_{\rm kin}$, exploiting the useful general property of any linearly varying mechanical system, namely  that its total mean energy is evenly divided between kinetic and potential energy. We thus have simply $E_{\rm kin}=\frac{1}{2}E_{\rm elastic}$, where the latter quantity was found in the previous section. We start from the general expression
\begin{equation}
E_{\rm kin}= \int_0^a \frac{1}{2}\rho_mV^2 2\pi\rho d\rho = \pi \rho_mA^2\sin^2\omega t\int_0^{ka}xJ_1^2(x)dx, \label{45}
\end{equation}
take the time average over a period, and make use of the following formula \cite{NIST},
\begin{equation}
\int x J_\mu^2(x)dx= \frac{1}{2}x^2[ J_\mu^2(x)-J_{\mu-1}(x)J_{\mu+1}(x)] \label{46}
\end{equation}
to get
\begin{equation}
E_{\rm kin}=\frac{1}{4}\pi \rho_mA^2(ka)^2|J_0(ka)J_2(ka)|. \label{47}
\end{equation}
Since this is equal to one half of the total energy $E_{\rm elastic}$ given in Eq.~(\ref{36}), we have now the opportunity to  determine the value of the constant $A$,
\begin{equation}A= \frac{w_0 \chi(0)}{\rho_m u ka}\Big| \frac{\int_0^{\tau_{\rm max}}F(\tau)d\tau}{J_0(ka)J_2(ka)} \Big|^{1/2}. \label{48}
\end{equation}
Inserting $w_0=2~\mu$m, \, $\chi(0)=4.9\times 10^5~$Pa, \, $J_0(ka)=-0.40, \, J_2(ka)= +0.40$ for $ka=3.83$, and adopting the same value 0.80  for the $\tau$ integral as above, we obtain
\begin{equation}
A= 3.8\times 10^{-7}~{\rm m^2 s}^{-1}. \label{49}
\end{equation}
The radial fluid velocity follows from Eq.~(\ref{42}). As an example, for an intermediate point $k\rho=1$ we find $V=(0.64\, \sin \omega t)~{\rm cm\, s}^{-1}.$  The fluid velocities are moderate. Although the ratio between the angular frequency and the wave number for the fluid wave is fixed by the velocity of sound, $\omega/k= u$, the magnitude of $V$ is much smaller than $u$ because of the weakness of the velocity potential $\Phi$.

It ought to be recalled that, instead of using the expression (\ref{47}), one might equally well calculate $E_{\rm kin}$ directly from Eq.~(\ref{36}),
\begin{equation}
E_{\rm kin}= \frac{\pi w_0^2}{4\rho_mu^2}\,\chi^2(0)\int_0^{\tau_{\rm max}}F(\tau)d\tau. \label{50}
\end{equation}
Turn now to the pressure. We will be concerned with the dynamic pressure $p'= -\rho_m\partial_t\Phi$ only. (The physiological range for blood pressure is between 80 and 120 mm Hg, where 1 mm Hg = 133 Pa, and plays in our context the role of the background pressure $p_0$.) For simplicity we omit hereafter the prime on $p$. From Eq.~(\ref{41a}) we have thus
\begin{equation}
p(\rho, t)=\rho_mA\omega J_0(k\rho)\cos \omega t. \label{51}
\end{equation}
As  $J_0(ka)$ is negative, the pressure $p(a)$ on the wall becomes positive in the half-period $\pi/2 < \omega t < 3\pi/2$. It is this pressure that we wish to maximize.  To make the effect appreciable, it is clear that one has to make use of a succession of short pulses. The  maximum case is when one  pulse is sent through the system each oscillation period, during the part of the cycle where the fluid is moving inwards, corresponding to $\pi < \omega t < 2\pi$ above.  Then each new pulse serves to increase the inward velocity. Physically,  this is a parametric resonance effect.

Ideally, we shall assume that by sending in  $N$ pulses per second, the pressure becomes enhanced by a factor $N$. Moreover, as another approximation we shall confine ourselves to calculating the root-mean-square (rms) pressure, called $\langle p\rangle_{\rm rms}$. The mean  pressure at the wall becomes accordingly
\begin{equation}
\langle p(a)\rangle_{\rm rms}= \frac{N\rho_m}{\sqrt{2}}A\omega |J_0(ka)|. \label{52}
\end{equation}
To obtain information about what influence   this pressure has on the wall, one has to take into account its  Young's modulus. Before entering into this, we shall  estimate the influence from viscosity on the oscillating fluid system.

\section{Viscous damping of the oscillating fluid}

The two viscosity coefficients of the fluid - shear viscosity  $\eta$ and bulk viscosity $\zeta$ - cause the oscillating motion to be damped out. For simplicity, we consider the  oscillating stationary wave picture from the previous section as our starting point, and will neglect the bulk viscosity as its magnitude seems to be largely unknown for  blood.  Measured values of $\eta$ lie roughly in the interval  from 2 to 5~$\times 10^{-3}~$Pa\,s. For definiteness, we adopt the value
\begin{equation}
\eta = 2\times 10^{-3}~\rm{Pa\,s}. \label{53}
\end{equation}
Consider now the time derivative $\dot{E}_{\rm kin}$ of the kinetic energy \cite{landau59},
\begin{equation}
\dot{E}_{\rm kin} = -\frac{1}{2}\eta \int [\partial_kV_i+\partial_iV_k -\frac{2}{3}\delta_{ik}{\bf (\nabla \cdot V)}^2]^2 d^3r. \label{54}
\end{equation}
We may calculate this expression by first employing Cartesian coordinates, as given in the expression above, and thereafter transfer the result to cylindrical coordinates. If we assume a one-dimensional harmonic motion along the $x$ axis, with velocity $V= V(x,t)$, we obtain after some manipulations
\begin{equation}
 \dot{E}_{\rm kin} = -\frac{4}{3}\eta \int (\partial_xV)^2 d^3r. \label{55}
 \end{equation}
Now replace $(\partial_xV)$ with ${\bf \nabla \cdot V}= \frac{1}{\rho}\frac{d}{d\rho}(\rho V)$ in cylindrical coordinates, and insert $V(\rho,t)$ from Eq.~(\ref{42}) to calculate
\begin{equation}
\dot{E}_{\rm kin} = -(\frac{2}{3}\eta) \pi k^2A^2\int_0^{ka} \frac{dx}{x}\left[ \frac{d}{dx}(xJ_1(x))\right]^2, \label{56}
\end{equation}
where we have averaged over one period. Now make use of Eqs.~10.6(ii) and 10.22.7 in Ref.~\cite{NIST},
\begin{equation}
\left(\frac{1}{x}\frac{d}{dx}\right)^k(x^\nu J_\nu (x))= x^{\nu-k}J_{\nu-k}(x), \label{57}
\end{equation}
\begin{equation}
\int x^{2\mu+1}J_\mu^2 (x)dx= \frac{x^{2\mu+2}}{2(2\mu+1)}
[ J_\mu^2(x) + J_{\mu+1}^2(x)], \quad 2\mu \neq -1, \label{58}
\end{equation}
to get
\begin{equation}
\dot{E}_{\rm kin} = -\frac{2}{3}\eta \pi (kA)^2 (ka)^2 J_0^2(ka). \label{59}
\end{equation}
We want to find how much of the initial kinetic energy $E_{\rm kin}$ becomes dissipated into heat per second. Calling the fraction $\gamma$ we have, using  the expression (\ref{47}) for $E_{\rm kin}$,
\begin{equation}
\gamma =\Big|\frac{ \dot{E}_{\rm kin}}{E_{\rm kin}}\Big| = \frac{8\eta}{3}\, \frac{k^2}{\rho_m}\, \Big|\frac{J_0(ka)}{J_2(ka)}\Big|. \label{60}
\end{equation}
With $\eta$ as given above, and with $a= 10^{-4}~$m, $ka=3.83$, this gives
\begin{equation}
\gamma= 7.8\times 10^3~{ \rm s}^{-1}. \label{61}
\end{equation}
This large number may at first appear to be a  difficulty for the present theory, but one needs here to consider how the numbers appear in the physical process: As observed above,  the maximum pressure on the symmetry axis is established  a very short time after the onset of the laser pulse, $\tau \sim 1$ or $t \sim 1~$ ns. The fluid rebounds immediately, and the return  wave propagates with velocity $u$ and reaches the wall in about 100 ns. If we assume as a simple model that the decay of kinetic energy occurs exponentially, $E_{\rm kin} \propto e^{-\gamma t}$, we see  that the  kinetic energy has by then diminished only by the fraction
\begin{equation}
\Big| \frac{\Delta E_{\rm kin}(100~{\rm ns})}{ E_{\rm kin}}\Big| \sim 0.1~\%. \label{62}
\end{equation}
There is thus ample time for the wave to rebound and strike upon the wall, before the viscosity begins to be appreciable.  Actually, the viscosity is a helpful factor here, because in the long run it reduces the amplitude of the contracting phases of the wall. This was in fact the physical idea behind the introduction of the rms pressure in Eq.~(\ref{52}). Namely, for
 an ideal nonviscous fluid a pulsating force on the wall without any damping would have given rise to an oscillating behavior of the wall only, about its equilibrium position, and thus not give a net outward force.

\section{Mean deformation of the wall.  }

Now return to the expression (\ref{52}) for the mean pressure on the artery wall, having seen that it is roughly justified to omit   the influence from  viscosity.  The maximum case, as mentioned above, is that one pulse is sent in synchronously with the natural oscillations of the fluid.   To estimate how much the wall becomes expanded by the repetitive bouncing forces, we must know the approximate value of Young's modulus $Y$ for the arteries. For most condensed bodies, $Y$ is known to lie in the GPa region. In the present case, however, the actual values are considerably smaller, in the MPa region.  Taking the internal mammary artery as an example, we may use data from Refs.~\cite{camasao21,stekelenburg09} to choose
\begin{equation}
Y= 10~{\rm MPa}. \label{63}
\end{equation}
Thus, using the expression (\ref{52}) for the mean pressure $ \langle p(a)\rangle_{\rm rms}$ we may write
\begin{equation}
\frac{N\rho_m}{\sqrt{2}}A\omega |J_0(ka)| = Y\, \frac{\Delta a}{a}, \label{64}
\end{equation}
where $\Delta a$ means a small or  moderate expansion of the radius $a$. Insertion of the number given above leads to the relationship
\begin{equation}
N \sim 10^3\, \frac{\Delta a}{a}. \label{65}
\end{equation}
This is actually a promising number  in our context. It means, for instance, that in order to make the radius to expand by a fraction of 0.1 we need a moderate number of about 100 successive laser pulses per second. This is much less than the theoretical optical limit which is one pulse per fluid oscillation frequency, which is in the MHz region as we have seen (Eq.~(\ref{44})). At least in principle, the effect seems to be  appreciable.

\section{Summary}

We have in this paper gone through the following items, and can summarize as follows:

\noindent 1) The basic idea behind  the above  analysis was to investigate whether a rapid succession of short laser pulses sent longitudinally through a long cylindrical tube of radius $a$ is able to expand the radius to some extent, via   the electrostrictive effect. The possibility of enlargement is of great interest in medical applications, and for that reason we used in our analysis a typical radius of $a= 100~\mu$m for a small blood artery. There may evidently be other technological applications also, and in most cases one would find it natural to assume larger dimensions than those used here.

\noindent 2) In Sec.~III we used Green function methods to describe  the build-up  of an inner hydrodynamic pressure in the central region, caused by one single laser pulse. The important scale in this initial part of the process is given by the velocity $u$ of sound. It   turns out that the pressure is established very quickly, when the nondimensional time parameter  $\tau$ defined in Eq.~(\ref{23}) is of order unity.

\noindent 3) In Sec.~IV we analyzed the  hydrodynamical problem with transverse oscillations in the tube after the laser pulse had left, and derived an expression for  the kinetic energy per unit length, as well as the rms pressure on the wall.

\noindent 4) Shear viscosity in the fluid was taken into account in Sec.~V. The effect of viscosity is to damp the hydrodynamic energy in general, although we found that the rebound of the fluid after the initial compression takes so short a time that the damping effectively does not come into play.

\noindent 5) As shown in Eq.~(\ref{65}), the number $N$ of successive laser pulses necessary to make an appreciable enlargement of the tube is moderate. So far, the results are promising.

\noindent 6) What can be said about future developments of this idea? Obviously, our analysis above is crude, and we should point out that heating effects have not been discussed at all. To the author it seems most natural that the proposed effect could be tested experimentally, preferably under larger geometric dimensions that those assumed in the biomedical case above. It is clear that the effect physically exists; what is unclear is its magnitude in practice.


\begin{thebibliography}{99}

\bibitem{specialissue} "Symmetry and Symmetry-Breaking in Fluid Dynamics", Special Issue of Symmetry, edited by A. Herczynski and R. Zenit (2023).

\bibitem{olsman21} M. Olsman, M. M{\"u}hlenpfordt, E. B. Olsen, S. H. Torp, S. Kotopoulis, C. J. F. Rijcken, Q. Hu, M. Thewissen, S. Snipstad, and C. de Lange Davies, Acoustic cluster therapy (ACT) enhances acumulation of polymeric micelles in the murine brain, J. Controlled Release {\bf 337}, 285 (2021).

\bibitem{poon22} C. Poon, M. M{\"u}hlenpfordt, M. Olsman, S. Kotopoulis, C. de Lange Davies, and K. Hynynen, Real-time intravital multiphoton microscopy to visualize focused ultrasound and microbubble treatments to increase blood-brain permeability, J. Vis. Exp. (180), e62235, doi:10.3791/62235 (2022).

\bibitem{bellettato18} C. M. Bellettato and M. Scarpa, Possible strategies to cross the blood-brain barrier, Italian J. Pediatrics {\bf 44} (Suppl. 2), 131 (2018).

\bibitem{liu22} D. Liu, X. Liu, Z. Chen, Z. Zuo, X. Tang, Q. Huang and T. Arai, Magnetically driven soft continuum microrobot for intravascular operations in microscale, Cyborg and Bionic Systems {\bf 2022}, Article 9850832, doi.org/10.34133/2022/9850832.

\bibitem{song21} P. Song, H. Fu, Y. Wang, C. Chen, P. Ou, R. T. Rashid, S. Duan, J. Song, Z. Mi and X. Liu, A microfluidic field-effect transistor biosensor with rolled-up indium nitride microtubes, Biosensors and Bioelectronics {\bf 190}, 113264 (2021).

\bibitem{lipsman18} N. Lipsman, Y. Meng,A. J. Bethune, Y. Huang, B. Lam, M. Masellis, N. Herrmann, C. Heyn, I. Aubert, A. Boutet, G. S. Smith, K. Hynynen and S. E. Black, Blood-brain opening in Alzheimer's disease using MR-guided focused ultrasound, Nature Comm. {\bf 9}, 2336 (2018).






\bibitem{nelson79} D. F. Nelson, {\it Electric, Optic, and Acoustic Interaction in Dielectrics} (Wiley, New York, 1979).

\bibitem{brevik79} I. Brevik, Experiments in phenomenological electrodynamics and the electromagnetic energy-momentum tensor, Phys. Reports {\bf 52}, 133 (1979).
 \bibitem{rasaiah81} J. C. Rasaiah, D. J. Ishister and G. Stell, Nonlinear effects in polar fluids: A molecular theory of electrostriction, J. Chem. Phys. {\bf 75}, 4707 (1981).
 \bibitem{carnie82}S. L. Carnie and G. Stell, Electrostriction and dielectric saturation in a polar fluid, J. Chem. Phys. {\bf 77}, 1017 (1982).
\bibitem{rasaiah82} J. C. Rasaiah, Electrostriction and the dielectrc constant of a simple polar fluid, J. Chem. Phys. {\bf 77}, 5710 (1982).
 \bibitem{brevik82}I. Brevik, Fluids in electric and magnetic fields: Pressure variation and stability, Can. J. Phys. {\bf 60}, 449 (1982).
 \bibitem{zimmerli99} G. A. Zimmerli, R. A. Wilkinson, R. A. Ferrell and M. R. Moldover, Electrostriction of near-critical fluid in microgravity, Phys. Rev. E {\bf 59}, 5862 (1999).
\bibitem{hallanger05}A. Hallanger, I. Brevik, S. Haaland and R. Sollie, Nonlinear deformations of liquid-liquid interfaces induced by electromagnetic radiation, Phys. Rev. E {\bf 71}, 056601 (2005).
\bibitem{birkeland08} O. J. Birkeland and I. Brevik, Nonlinear laser-induced deformations of liquid-liquid interfaces: An optical fiber model, Phys. Rev. E {\bf 78}, 066314 (2008).
\bibitem{brasselet08} E. Brasselet, R. Wunenburger and J.-P. Delville, Liquid optical fibers with a multistable core actuated by light radiation pressure, Phys. Rev. Lett. {\bf 101}, 014501 (2008).
\bibitem{delville09} J.-P. Delville, M. R. de Sait Vincent, R. D. Schroll, H. Chraibi, B. Issenmann, R. Wunenburger, D. Lasseux, W. W. Zhang  and E. Brasselet, Laser microfluidics: Fluid actuation by light, J. Opt. A: Pure Appl. Opt. {\bf 11}, 034015 (2009).
\bibitem{wunenburger11} R. Wunenburger, B. Issenmann, E. Brasselet, C. Loussert, V. Hourtane and J.-P. Delville, Fluid flows driven by light scattering, J. Fluid Mech. {\bf 666}, 273  (2011).





\bibitem{ellingsen11} S. {\AA.} Ellingsen and I. Brevik, Electrostrictive fluid pressure from a laser beam, Phys. Fluids {\bf 23}, 096101 (2011).

\bibitem{ellingsen12} S. {\AA.} Ellingsen and I. Brevik,  Electrostrictive counterforce on fluid microdroplet in short laser pulse, Opt. Lett. {\bf 37}, 1928 (2012).

\bibitem{astrath14} N. G. C. Astrath, L. C. Malacarne, M. L. Baesso, G. V. B. Lukasievicz and S. E. Bialkowski, Unravelling the effects of radiation forces in water, Nature Communications {\bf 5}, 4363 (2014).

\bibitem{astrath22} N. G. C. Astrath, G. A. S. Flizikowski, B. Anghinoni, L. C. Malacarne, M. L. Baesso, T. Po\^{z}ar, M. Partanen, I. Brevik, D. Razansky and S. E. Bialkowki, Light: Science \& Applications  {\bf 11}, 103 (2022).

\bibitem{astrath23} N. G. C. Astrath, B. Anghinoni, G. A. S. Flizikowski, V. S. Zanuto,  L. C. Malacarne, M. L. Baesso,  T. Po\^{z}ar and D. Razansky, The role of electrostriction in the generation of acoustic waves by optical forces in water, preprint, 2023.

\bibitem{stratton41}
J. A. Stratton, {\it Electromagnetic Theory}  (McGraw-Hill, New York, 1941).

\bibitem{jackson99} J. D. Jackson, {\it Classical Electrodynamics}, 3rd ed. (Wiley, New York,1998).

\bibitem{camasao21} D. B. Camasao and D. Mantovani, The mechanical characterization of blood vessels and their sustitutes in the continuous quest for physiological-relevant  performances. A critical review, Materials Today Bio {\bf 10}, 100106 (2021).

\bibitem{ashkin73} A. Ashkin and J. M. Dziedzic, Radiation pressure on a free liquid surface,  Phys. Rev. Lett. {\bf 30}, 179 (1973).

\bibitem{NIST} {\it NIST Handbook of Mathematical Functions}, edited by F. W. J. Olver {\it et al.} (Cambridge University Press, Cambridge, 2010), Eq.~10.22.5.

\bibitem{landau59} L. D.Landau and E. M. Lifshitz, {\it Fluid Mechanics}  (Pergamon Press, Oxford, 1959).

\bibitem{stekelenburg09}  M. Stekelenburg, M. C. Rutten, L. H. Snoeckx and F. P. Baaljens, Dynamic straining combined with fibrin gel cell seeding improves strength of tissue-engineered small-diameter vascular grafts, Tissue Eng. {\bf 15}, 1081 (2009).


\end{thebibliography}
\end{document}